\documentclass[pra,aps,twocolumn,epsfig,superscriptaddress,showpacs]{revtex4}
\usepackage{amsmath}
\usepackage{amssymb}
\usepackage{graphicx}
\usepackage{dcolumn}
\usepackage{bm}
\usepackage[colorlinks=false,dvipdfm]{hyperref} 
\usepackage{setspace}
\usepackage{amsfonts}
\usepackage{rotating}
\usepackage{makeidx}
\usepackage{amsthm}

\newcommand{\tr}{\mbox{Tr}}

\begin{document}


\title{Requirement of Dissonance in Assisted Optimal State Discrimination \footnote{accepted by Scientific Reports}}

\author{Fu-Lin Zhang\footnote{Correspondence to: flzhang@tju.edu.cn}}
\affiliation{Physics Department, School
of Science, Tianjin University, Tianjin 300072, China}

\author{Jing-Ling Chen\footnote{Correspondence to: cqtchenj@nus.edu.sg}}
\affiliation{Theoretical Physics Division, Chern Institute of
Mathematics, Nankai University, Tianjin, 300071, China}
\affiliation{Centre for Quantum Technologies, National University of
Singapore, 3 Science Drive 2, Singapore 117543}

\author{L.~C.~Kwek}
\affiliation{Centre for Quantum Technologies, National University of
Singapore, 3 Science Drive 2, Singapore 117543}
 \affiliation{National Institute of Education,
 Nanyang Technological University, 1 Nanyang Walk, Singapore 637616}
 \affiliation{Institute of Advanced Studies, Nanyang Technological University,
60 Nanyang View, Singapore 639673}

\author{Vlatko Vedral}
\affiliation{Centre for Quantum Technologies, National University of
Singapore, 3 Science Drive 2, Singapore 117543}
\affiliation{Clarendon Laboratory, University of Oxford, Parks Road,
Oxford OX1 3PU, United Kingdom} \affiliation{Department of Physics,
National University of Singapore, 2 Science Drive 3, Singapore
117542}

\date{\today}

\begin{abstract}
\textbf{A fundamental problem in quantum information is to explore what
kind of quantum correlations is responsible for successful
completion of a quantum information procedure.
Here we study the roles of entanglement, discord, and dissonance needed for optimal
quantum state discrimination when the latter is assisted with an auxiliary system.
In such process, we present a more general joint unitary transformation than the existing results.
The quantum entanglement
between a principal qubit and an ancilla is found to be completely unnecessary, as it can be set to zero in the
arbitrary case  by adjusting the parameters in the general unitary without
affecting the success probability.
 This result also shows that it is quantum
dissonance that plays as a key role in assisted optimal state
discrimination and not quantum entanglement. A necessary criterion
for the necessity of quantum dissonance based on the linear entropy is also
presented.}
\end{abstract}



\maketitle






An important distinctive feature of quantum mechanics is that quantum coherent
superposition can lead to quantum correlations in composite
quantum systems like quantum entanglement \cite{Horodecki09}, Bell
nonlocality \cite{Bell} and quantum discord
\cite{PhysRevLett.88.017901,henderson2001classical}. Quantum
entanglement has been extensively studied from various perspectives,
and it has served as a useful resource for demonstrating the
superiority of quantum information processing. For instance,
entangled quantum states are regarded as key resources for some quantum
information tasks, such as teleportation, superdense coding and
quantum cryptography \cite{key}.

In contrast to quantum entanglement, quantum discord measures the amount of nonclassical correlations between two subsystems of a bipartite quantum system. A recent report regarding the
deterministic quantum computation with one qubit (DQC1)
\cite{lanyon2008experimental,datta2008quantum} demonstrates that a
quantum algorithm to determine the trace of a unitary matrix can
surpass the performance of the corresponding classical algorithm in terms of computational speedup
even in the absence of quantum entanglement between the the control
qubit and a completely mixed state. However, the quantum discord is never zero.
This result is somewhat surprising and it
has engendered much interest in quantum discord in recent years. In particular, it has led to further
studies  on the relation of quantum discord with other measures of correlations.  Moreover, it
has been shown that it is possible to formulate an operational
interpretation in the context of a quantum state merging protocol
\cite{cavalcanti2011operational,madhok2011interpreting} where it
can be regarded as the amount of entanglement generated in an activation protocol
\cite{piani2011all} or in a measurement process
\cite{streltsov2011linking}.  Also, a unified view of quantum
correlations based on the relative entropy \cite{modi2010unified}  introduces a
new measure called quantum dissonance which can be regarded as the nonclassical
correlations in which quantum entanglement has been totally excluded. For a
separable state (with zero entanglement), its quantum dissonance is
exactly equal to its discord.


It is always interesting to uncover non-trivial  roles of
nonclassical correlations in quantum information processing.
The quantum algorithm in DQC1 has been widely regarded as
the first example for which quantum discord, rather than quantum entanglement,
plays  a key role in the computational process. Moreover, a careful consideration of the natural
bipartite split between the control qubit and the input state reveals that the
quantum discord is nothing but the quantum dissonance of the system. This simple observation naturally leads to
an interesting question: Can quantum dissonance serve as a
similar key resource in some quantum information tasks? The affirmative
answer was shown in an interesting piece of work  by
Roa, Retamal and Alid-Vaccarezza \cite{roa2011dissonance} where the roles of
entanglement, discord, and dissonance needed for performing
unambiguous quantum state discrimination assisted by an auxiliary
qubit  \cite{john1996mathematical,POVM2008} was studied.  
This protocol for \emph{assisted optimal state discrimination} (AOSD)
in general requires both quantum entanglement and discord. However, for the case in which there
exist equal \emph{a priori} probabilities, the entanglement of the state of system-ancilla
qubits is absent even though its discord is nonzero, and hence the
unambiguous state discrimination protocol is implemented
successfully only with quantum dissonance.  This
protocol therefore provides an example for
which dissonance, and not entanglement, plays as a key role  in a quantum information processing task.

In this work, we show more generally that quantum entanglement is
not even necessary for  AOSD.  Moreover, we look at the roles of
correlations in the AOSD under the most general settings by
considering a generic AOSD protocol. We also show that only
dissonance in general is required for AOSD and quantum entanglement
is never needed.


\section{Results}

\begin{figure}
\includegraphics[width=8cm]{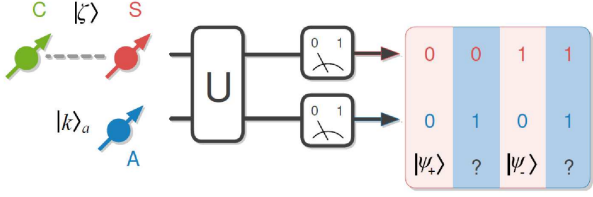} \\
\caption{ \textbf{The General AOSD Protocol Illustration.} Alice and Bob share a pure entangled state $|\zeta\rangle$ of qubits $S$ and $C$. To discriminate the two states
$|\psi_+\rangle$ or $|\psi_-\rangle$ of $S$, Alice performs a joint
unitary transformation $\mathcal {U}$ between qubits $S$ and $A$, followed by
two independent von Neumann measurements on the two qubits. Her
state discrimination is successful if the outcome of $A$ is $0$, but
unsuccessful if outcome $1$.} \label{fig1}
\end{figure}

\noindent \textbf{The General AOSD Protocol.} Suppose Alice and Bob
share an entangled two-qubit state
$|\zeta\rangle=\sqrt{p_+}|\psi_+\rangle|0\rangle_c+\sqrt{p_-}|\psi_-\rangle|1\rangle_c$
(see Fig.~\ref{fig1}), where $p_{\pm} \in [0,1]$ and $p_++p_-=1$,
 $|\psi_{\pm}\rangle$ are two nonorthogonal states of the qubit of Alice (system qubit $S$),
and $\{|0\rangle_c,|1\rangle_c \}$ are the orthonormal bases for the one of Bob (qubit $C$).
The reduced state of system qubit  $ \rho=p_+ |\psi_+\rangle \langle\psi_+|+p_-
|\psi_-\rangle \langle\psi_-|$ is a realization of the
model in \cite{roa2011dissonance} in which a qubit is
prepared in the two nonorthogonal states $|\psi_{\pm}\rangle$ with \emph{a priori} probabilities $p_{\pm}$.
 To discriminate the two states
$|\psi_+\rangle$ or $|\psi_-\rangle$ unambiguously,
the system is coupled to an auxiliary qubit $A$, prepared in a known
initial pure state $|k\rangle_a$. Under a joint unitary
transformation $\mathcal {U}$ between the system and the ancilla,
one obtains
\begin{subequations}\label{Utrans}
\begin{align}
\mathcal {U}\;|\psi_+\rangle|k\rangle_a=\sqrt{1-|\alpha_+|^2}\;|0\rangle|0\rangle_a+\alpha_+|\Phi\rangle|1\rangle_a, \\
\mathcal
{U}\;|\psi_-\rangle|k\rangle_a=\sqrt{1-|\alpha_-|^2}\;|1\rangle|0\rangle_a+\alpha_-|\Phi\rangle|1\rangle_a,
\end{align}
\end{subequations}
where $|\Phi\rangle=\cos \beta |0 \rangle+ \sin \beta e^{i \delta}
|1\rangle$, $\{|0\rangle,|1\rangle\}$ and
$\{|0\rangle_a,|1\rangle_a\}$ are the bases for the system and the
ancilla, respectively.
The probability amplitudes $\alpha_+$ and $\alpha_-$ satisfy
$\alpha_+^\ast\alpha_-=\alpha$, where
$\alpha=\langle\psi_+|\psi_-\rangle=|\alpha|e^{i\theta}$ is the
\emph{priori} overlap between the two nonorthogonal states. The
unitary transformation can be constructed by performing an operation $\mathcal{W}=(|+\rangle \langle 0 |+|-\rangle \langle 1 |) \otimes |0\rangle_a\langle0| + (|\Phi\rangle \langle 0 |+|\bar{\Phi}\rangle \langle1 |) \otimes |1\rangle_a\langle1|$ on the original one in Ref. \cite{roa2011dissonance}, where $|\pm\rangle=(|0\rangle \pm |1\rangle)/\sqrt{2}$ and $|\bar{\Phi}\rangle=\sin \beta |0 \rangle- \cos \beta e^{i \delta}
|1\rangle$. It has the form as
\begin{eqnarray}
\mathcal{U}=\frac{1}{1-|\alpha|^2}\biggr[
\left(\sqrt{1-|\alpha_+|^2}\;|0\rangle|0\rangle_a+\alpha_+|\Phi\rangle|1\rangle_a
\right)\langle \tilde{\psi}_+| _a\langle k| \nonumber \\
 +\left(\sqrt{1-|\alpha_-|^2}\;|1\rangle|0\rangle_a+\alpha_-|\Phi\rangle|1\rangle_a
\right) \langle \tilde{\psi}_-| _a\langle k|  \biggr] +
\mathcal{V}, \nonumber
\end{eqnarray}
where $|\tilde{\psi}_{\pm}\rangle= |\psi_{\pm}\rangle - |\psi_{\mp}\rangle
\langle \psi_{\mp} |\psi_{\pm}\rangle $ are the components of $
|\psi_{\pm}\rangle$ orthogonal to $|\psi_{\mp}\rangle$, and $\mathcal{V}= |\Upsilon_+\rangle \langle0| _a\langle \bar{k}|+|\Upsilon_-\rangle \langle1| _a\langle \bar{k}|$, with $ |\Upsilon_{\pm}\rangle$ being
two arbitrary states orthogonal to the right hands of  Eq.
(\ref{Utrans}) and $\langle\Upsilon_-|\Upsilon_+\rangle=0$, and
$_a\langle \bar{k}|k\rangle_a=0 $.
Obviously, only the terms with $\langle \tilde{\psi}_{\pm}| _a\langle k|$
have effect on the initial
state $|\psi_{\pm}\rangle |k\rangle_a  $.


The state of the system-ancilla qubits is given by
\begin{eqnarray}\label{rho}
\rho_{SA} & = & p_+ \;\mathcal {U}\left( |\psi_+\rangle\langle\psi_+|\otimes |k\rangle_a \langle k|\right) \mathcal {U}^\dag\nonumber\\
&  & + p_-\;\mathcal {U}\left( |\psi_-\rangle \langle\psi_-|\otimes
|k\rangle_a \langle k|\right) \mathcal {U}^\dag,
\end{eqnarray}
which depends on $\beta$ and $\delta$, and it is generally not
equivalent to the corresponding one in \cite{roa2011dissonance}
under local unitary transformations unless $|\Phi\rangle=|+\rangle$.
The state discrimination is successful
if the ancilla collapses to $|0\rangle_a$.  This occurs with success
probability given by
 \begin{eqnarray}\label{prob}
P_{\rm suc}&=&{\rm Tr}[(\openone_s\otimes
|0\rangle_a\langle0|)\rho_{SA}
]\nonumber\\
&=&p_+ (1-|\alpha_+|^2)+p_- (1-|\alpha_-|^2),
\end{eqnarray}
where $\openone_s$ is the unit matrix for the system qubit.
Without
loss of generality, let us assume that
$p_+\leq p_-$ and denote $\bar{\alpha}=\sqrt{p_+/p_-}$. The analysis
of the optimal success probability can be divided into two cases:
(i) $|\alpha| < \bar{\alpha}$, $P_{\rm suc}$ is attained for
$|\alpha_+|=\sqrt[4]{p_{-}/p_{+}}\sqrt{|\alpha|}$; (ii) $
\bar{\alpha}\leq |\alpha|\leq 1$, $P_{\rm suc}$ is attained for
$|\alpha_+|=1$ (or equivalently $|\alpha_-|=|\alpha|$). One has
\begin{subequations}\label{twocase}
\begin{align}
P_{\rm suc, max} = 1-2 \sqrt{p_+ p_-} |\alpha|, \;\;\;\; {\rm for \;\; case (i)},\label{case1}&&\\
P_{\rm suc, max} = (1-|\alpha|^2) p_-, \;\;\;\;\;\;\;\; {\rm for
\;\; case (ii)}.\label{case2}&&
\end{align}
\end{subequations}
Before proceeding further to explore the roles of correlations in the AOSD,
we make the following remarks.

\medskip

\noindent \emph{Remark 1.} State discrimination of a subsystem in a reduced
mixed state has practical interest in conclusive quantum
teleportation where the resource is not prepared in a maximally
entangled state (see Refs. \cite{Horo,Kim1,Kim2}). In the conclusive
teleportation protocol, the sender Alice possesses an
arbitrary one-qubit state $|{\varphi}\rangle_{\rm
Alice}=a|0\rangle+b|1\rangle$, and she shares a non-maximally
entangled state $|\Psi_+(\theta)\rangle=\cos\theta|00\rangle
+\sin\theta|11\rangle$ with the receiver Bob. Under the protocol,
one has
\begin{eqnarray}
|\Psi_{\rm tel}\rangle&=&|\varphi\rangle_{\rm
Alice} \otimes |\Psi_+(\theta)\rangle\nonumber \\
&=&\frac{1}{2}\biggr\{|\Psi_+(\theta)\rangle\otimes
|\varphi\rangle_{\rm Bob}
+ |\Psi_-(\theta)\rangle\otimes \sigma_z|\varphi\rangle_{\rm Bob} \nonumber \\
&&+   |\Phi_+(\theta)\rangle\otimes \sigma_x|\varphi\rangle_{\rm Bob}
+ |\Phi_-(\theta)\rangle\otimes (-i\sigma_y)|\varphi\rangle_{\rm
Bob} \biggr\},\nonumber
\end{eqnarray}
where $|\Psi_{\pm}(\theta)\rangle=\cos\theta|00\rangle \pm
\sin\theta|11\rangle$,
$|\Phi_{\pm}(\theta)\rangle=\sin\theta|01\rangle \pm
\cos\theta|10\rangle$, and $\sigma_x, \sigma_y,
\sigma_z$ are Pauli matrices. The concurrences \cite{Wootters98} of the
states $|\Psi_{\pm}(\theta)\rangle$ and $|\Phi_{\pm}(\theta)\rangle$
are all
equal to $\mathcal {C}=|\sin2\theta|$. The states
$|\Psi_{\pm}(\theta)\rangle$ are orthogonal to the states
$|\Phi_{\pm}(\theta)\rangle$, but
$\{|\Psi_+(\theta)\rangle,|\Psi_-(\theta)\rangle\}$ (or
$\{|\Phi_+(\theta)\rangle,|\Phi_-(\theta)\rangle\}$) are not
mutually orthogonal. To teleport the unknown state
$|{\varphi}\rangle_{\rm Alice}$ from Alice to Bob with perfect fidelity
(equals to 1), state discrimination
\cite{Horo,Kim1,Kim2} is generally required.  It should also be noted that only the maximally
entangled states (with $\theta=\pi/4$) can realize the perfect
teleportation with unit success probability.

\medskip

\noindent \emph{Remark 2.} Through quantum teleportation, we see that our model recover  the scheme in \cite{roa2011dissonance}, in which the principal qubit is randomly prepared in one of the two pure states $|\psi_+\rangle$ or $|\psi_-\rangle$. Let us conisder replacing the
entangled resource $|\Psi_+(\theta)\rangle$ by maximally entangled states randomly prepared with a probabilities as $\{p_1: |\Psi_+(\frac{\pi}{4})\rangle,\ p_2: |\Psi_-(\frac{\pi}{4})\rangle,\ p_3:|\Phi_+(\frac{\pi}{4})\rangle,\  p_4: |\Phi_-(\frac{\pi}{4})\rangle \}$.
Although they are all maximally entangled
states and each of them is a resource for perfect teleportation,
perfectly faithful teleportation cannot be realized in this case.
It can be shown that the fidelity of teleportation is the one
corresponding to the \emph{average state} \cite{Horo}
 \begin{eqnarray}\label{mixedstate}
\rho_{\rm res}&=&p_1  |\Psi_+(\frac{\pi}{4})\rangle \langle \Psi_+(\frac{\pi}{4})| + p_2  |\Psi_-(\frac{\pi}{4})\rangle \langle \Psi_-(\frac{\pi}{4})|  \nonumber \\
&\ & +p_3 |\Phi_+(\frac{\pi}{4})\rangle \langle
\Phi_+(\frac{\pi}{4})| + p_4 |\Phi_-(\frac{\pi}{4})\rangle \langle
\Phi_-(\frac{\pi}{4})|.\nonumber
\end{eqnarray}
Consequently, the amount of entanglement contributing to
teleportation is not just the average value of the entanglement which is
$p_1\mathcal {C}(|\Psi_+(\frac{\pi}{4})\rangle)+p_2\mathcal
{C}(|\Psi_-(\frac{\pi}{4})\rangle)+p_3\mathcal
{C}(|\Phi_+(\frac{\pi}{4})\rangle)+p_4\mathcal
{C}(|\Phi_-(\frac{\pi}{4})\rangle)$, but the entanglement of the
average state as $\mathcal {C} (\rho_{\rm res})$. Therefore the
amount of entanglement available depends crucially on the knowledge
of the entangled state. The amount of quantum entanglement that is needed for
the AOSD scheme  considered here, as well as the one in Ref.
\cite{roa2011dissonance}, refers to the entanglement of the average
state, $\mathcal {C} (\rho_{SA})$, and not to the average value of the
entanglement as $p_+ \mathcal {C}(\mathcal
{U}\;|\psi_+\rangle|k\rangle_a)+p_- \mathcal {C}(\mathcal
{U}\;|\psi_-\rangle|k\rangle_a)$.

We are now  ready to investigate the roles of correlations in
the AOSD. To this end, let us first  calculate the concurrence of
$\rho_{SA}$:
\begin{eqnarray}\label{con}
\mathcal {C}(\rho_{SA})&=& 2 \biggr[ \mathcal {Y}_+^2 \sin^2 \beta + \mathcal {Y}_-^2 \cos^2 \beta \nonumber \\
 && -2 \mathcal {Y}_+\mathcal {Y}_- \sin\beta \cos\beta
\cos(\theta+\delta)  \biggr]^{1/2},
\end{eqnarray}
with $\mathcal {Y}_\pm=\sqrt{1-|\alpha_\pm|^2}\;|\alpha_\pm| p_\pm$.
When $\beta=\pi/4$ and $\delta=0$, Eq. (\ref{con}) reverts to the
result in \cite{roa2011dissonance}.

Let us impose the constraint $\mathcal {C}(\rho_{SA})=0$ for any
$\alpha$, $\alpha_+$ and $p_+$. It is then easy to see that
\begin{eqnarray}\label{Czero}
\delta=-\theta, \ \ \
  \beta = \arctan (\mathcal{Y}_- / \mathcal{Y}_+).
\end{eqnarray}
Based on Eq. (\ref{Czero}), state (\ref{rho}) is a separable state
as
\begin{eqnarray}\label{rhosepa}
\rho_{SA}=|\eta_1 \rangle \langle \eta_1| \otimes
|0\rangle_a\langle0|+ |\Phi\rangle\langle\Phi|\otimes |
\eta_2\rangle_a \langle \eta_2 |,
\end{eqnarray}
where $|\eta_1 \rangle$ and $|\eta_2 \rangle_a$ are two unnormalized states as
\begin{eqnarray}
&&|\eta_1 \rangle =\frac{\sqrt{p_+ p_-}}{ \mathcal {Z}
}(\sqrt{1-|\alpha_+|^2} \alpha_- | 0\rangle
-\sqrt{1-|\alpha_-|^2} \alpha_+ |1\rangle), \nonumber \\
&&|\eta_2 \rangle_a = \frac{\sqrt{\mathcal {Y}_+^2+\mathcal
{Y}_-^2}}{\mathcal {Z}} \;| 0\rangle_a +\frac{\mathcal {Z}
\alpha_+}{|\alpha_+|}\;|1\rangle_a.
\end{eqnarray}
where $\mathcal {Z}=\sqrt{p_+ |\alpha_+|^2 + p_- |\alpha_-|^2}$.

Note that the state (\ref{rho}) has rank two, and  it is really the
reduced state of the following tripartite pure state
\begin{eqnarray}\label{3qubit}
|\Psi \rangle =\sqrt{p_+}\;(\mathcal {U}|\psi_+\rangle|k\rangle_a)|
0\rangle_c + \sqrt{p_-}\;(\mathcal {U}|\psi_-\rangle|k\rangle_a)|
1\rangle_c.
\end{eqnarray}
Its discord
can be derived analytically as $D(\rho_{SA})=S(\rho_A)-S(\rho_{SA})+E(\rho_{SC})=S(\rho_A)-S(\rho_{C})+E(\rho_{SC})$
using  the Koashi-Winter identity
\cite{koashi2004monogamy},
where $S(\rho)$ is the von Neumann entropy, $E(\rho_{SC})$ is the
entanglement of formation \cite{Wootters98} between the principal
system and the qubit $C$.  The explicit expression for the discord is
\begin{eqnarray}\label{discord1}
D(\rho_{SA})=\mathcal{H}(\tau_{A})-\mathcal{H}(\tau_{C})+\mathcal{H}(\tau_{SC}),
\end{eqnarray}
where $\mathcal{H}(x)=-\frac{1+\sqrt{1-x}}{2}\ln
\frac{1+\sqrt{1-x}}{2}-\frac{1-\sqrt{1-x}}{2}\ln
\frac{1-\sqrt{1-x}}{2}$,
$\tau_A$ is the tangle between $A$ and $SC$, $\tau_C$ is the tangle
between $C$ and $SA$, and $\tau_{SC}=\mathcal {C}^2(\rho_{SC})$ is
the concurrence between $S$ and $C$ in the state $\rho_{SC}$. One
can obtains
\begin{subequations}\label{threetau}
\begin{eqnarray}
&&\tau_{A}=\tau_{SA} +4p_+ p_- (|\alpha_+|^2+|\alpha_-|^2-2|\alpha|^2),\\
&&\tau_{C}=4 p_+ p_- (1-|\alpha|^2) ,\\
&&\tau_{SC}=\tau_S-\tau_{SA}-\tau(|\Psi\rangle),
\end{eqnarray}
\end{subequations}
with  $\tau_S$ the tangle between $S$ and $AC$, $\tau_{SA}=\mathcal
{C}^2(\rho_{SA})$, and $\tau(|\Psi\rangle)$ the three-tangle
\cite{coff}. The tangle between $S$ and $AC$ is given by
\begin{eqnarray}\label{taus}
\tau_S&= &4 \biggr\{(p_-|\alpha_-|^2 + p_+ |\alpha_+|^2) \nonumber \\
&& \times \left[p_-
(1-|\alpha_-|^2)\cos^2 \beta  + p_+ (1-|\alpha_+|^2)\sin^2 \beta \right] \nonumber \\
&& +p_+ p_- (1-|\alpha_+|^2)(1-|\alpha_-|^2) \biggr\},
\end{eqnarray}
and the three-tangle is
\begin{eqnarray}\label{tangle}
\tau(|\Psi\rangle)&=&
4 p_+ p_- \biggr|\sqrt{1-|\alpha_-|^2}\alpha_+\cos \beta  \nonumber \\
 && \ \ \ \ \  \ \ \ \ \   +\sqrt{1-|\alpha_+|^2}\alpha_-  \sin \beta e^{i \delta}\biggr|^2.
\end{eqnarray}

\begin{figure}
\includegraphics[width=8cm]{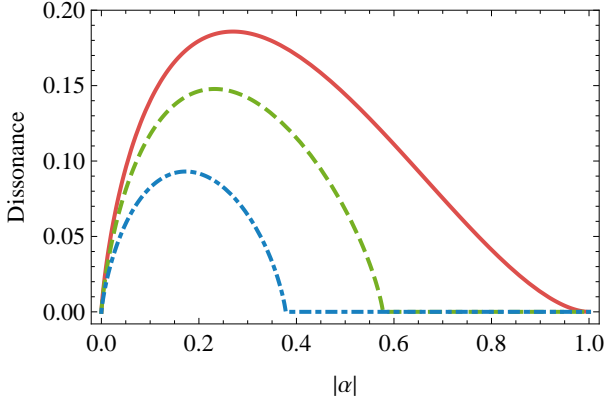} \\
 \caption{\textbf{Quantum dissonance in the AOSD.} We plot the dissonance versus $|\alpha|$, for $p_+=1/2$ (solid line), $1/4$
  (dashed line), and $1/8$ (dot-dashed line). Dissonance is greater than zero for case (i), and is zero for case
  (ii). The critical point for $D(\rho_{SA})=0$ occurs at
  $|\alpha|=\bar{\alpha}$.
} \label{fig2}
\end{figure}

\textbf{Dissonance for cases (i) and (ii).} For case (i), upon
the substitution $|\alpha_+|=\sqrt[4]{p_{-}/p_{+}}\sqrt{|\alpha|}$,
$p_-=1-p_+$, and Eqs.
(\ref{Czero})(\ref{threetau})(\ref{taus})(\ref{tangle}) into Eq.
(\ref{discord1}), one has the analytical expression for the
dissonance, which depends only on $|\alpha|$ and $p_+$. In Fig.
\ref{fig2}, we plot the curves of the dissonance versus $|\alpha|$
for $p_+=1/2, 1/4, 1/8$, respectively (see the curves with
$D(\rho_{SA})>0$). For case (ii), because $|\alpha_+|=1$, one has
$\beta=\pi/2$ and the state $\rho_{SA}$ is
\begin{eqnarray}\label{produ}
\rho_{SA}=|1\rangle\langle1|\otimes\rho_a,
\end{eqnarray}
with $\rho_a=p_+|1\rangle _a\langle 1|+p_-
|\mu\rangle_a\langle\mu|$,
$|\mu\rangle_a=\sqrt{1-|\alpha|^2}\;|0\rangle_a+\alpha_-e^{i\delta}|1\rangle_a$.
The state (\ref{produ}) is clearly a direct-product state hence
its dissonance is zero. In Fig. \ref{fig2}, for case (ii), we also
plot the curves of dissonance versus $|\alpha|$ for the same $p_+$'s
(see the curves with $D(\rho_{SA})=0$). Fig. \ref{fig2} shows
that dissonance is a key  ingredient for AOSD other than
entanglement for case (i), and that the classical state can accomplish
the task of AOSD for case (ii).


\begin{figure}
\includegraphics[width=8cm]{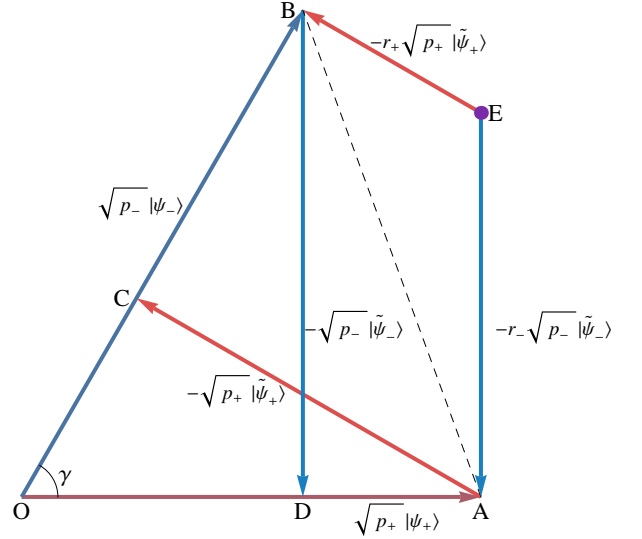} \\
 \caption{\textbf{Geometric picture for optimal success probability based on POVM strategy.}
 The sides $|OA|=\sqrt{p_+}$ , $|OB|=\sqrt{p_-}$, the angle $\gamma=\arccos |\alpha|$,
 and $AC \perp OB$, $BD \perp OA$, $EB \perp OB$, $EA \perp OA$. For
 $|\alpha|<\bar{\alpha}$, the point $E$ locates inside  of the angle $\angle
AOB$; for $|\alpha|\ge \bar{\alpha}$, the point $E$ coincides with
the point $B$ for $p_+ < p_-$ (or $A$ for  $p_+ > p_-$).}
 \label{fig3}
\end{figure}

\textbf{Geometric Picture.} It can be observed that
the optimal success probability $P_{\rm suc, max}$ in Eq.
(\ref{twocase}) can be analyzed  in two different regions: $|\alpha| <
\bar{\alpha}$ and $|\alpha| \geq \bar{\alpha}$. Here based on the
positive-operator-valued measure (POVM)
  strategy \cite{POVM2008}, we provide a geometric picture
of  $P_{\rm suc, max}$. Since the success probability, the
concurrence and the discord of state $\rho_{SA}$ under the
constraints in Eq. (\ref{Czero}) are all independent of the phase
$\theta$ of $\alpha$, one can simply set $\theta=0$, and
regard the states $|\psi_{\pm}\rangle$ as two unit vectors in
$\mathbb{R}^2$ with the angle $\gamma=\arccos |\alpha|$ between
them. The square roots of the \emph{a priori} probabilities, i.e.,
$\sqrt{p_+}$ and $\sqrt{p_-}$, behave like wave
amplitudes, and the effects of the coherence
can be seen from the states $|\zeta\rangle$ and
$|\Psi\rangle$. In Fig. \ref{fig3}, we plot two vectors
$\overrightarrow{OA}$ and $\overrightarrow{OB}$ with $\angle BOA
=\gamma$
 to denote $\sqrt{p_+}|\psi_{+}\rangle$ and $\sqrt{p_-}|\psi_{-}\rangle$, respectively.
The two POVM elements that identify the states
$\sqrt{p_\pm}|\psi_{\pm}\rangle$ can be implemented as
$\Pi_{\pm}=r_{\pm} |\tilde{\psi}_{\pm}\rangle \langle
\tilde{\psi}_{\pm} |/(\langle \tilde{\psi}_{\pm}
|\tilde{\psi}_{\pm}\rangle)$, with $r_\pm \ge 0$.
The vectors
$\overrightarrow{AC}$ and $\overrightarrow{BD}$  correspond to the
unnormalized states $|\tilde{\psi}_{\pm}\rangle$ with the
coefficients $-\sqrt{p_{\pm}}$. The third POVM element giving the
inconclusive result is $\Pi_0=\openone_s - \Pi_+-\Pi_-$. The
elements $\Pi_{\pm,0}$ are required to be positive -  this is a
constraint on the POVM strategy. Finally, the probability of successful
discrimination is $P_{POVM}=(r_+ p_+ + r_- p_-) (1-|\alpha|^2)$,
which is
\begin{eqnarray}\label{Ppovm}
P_{POVM}=(\overrightarrow{OA}-\overrightarrow{OB}) \cdot (r_- \overrightarrow{BD}- r_+ \overrightarrow{AC}).
\end{eqnarray}
When $|\alpha|< \bar{\alpha}$, the optimal $P_{POVM}$ is attained at
$r_{\pm} = (1-\cos \gamma\sqrt{p_{\mp} / p_{\pm}})/\sin^2 \gamma$.
The vectors $r_- \overrightarrow{BD}=\overrightarrow{EA}$ and
$r_+ \overrightarrow{AC}=\overrightarrow{EB}$,
 where $E$ is the intersection point of $AE$ and $BE$ (see Fig. \ref{fig3}).
The maximum value of $P_{POVM}$ is the square of $|AB|$, nanmely
$P_{POVM}=|AB|^2=1-2 \sqrt{p_+ p_-}|\alpha|$, which recovers Eq.
(\ref{case1}). When $|\alpha|= \bar{\alpha}$,  the point $E$
coincides with $B$ for $p_+ < p_-$ (or $A$ for  $p_+ > p_-$), for
the optimal $P_{POVM}$ one has $r_-=1$ (or $r_+=1$) and $P_{POVM}
=p_- (1-|\alpha|^2)$.
For $|\alpha|> \bar{\alpha}$ and $p_+ < p_-$, $E$ lies outside of
the angle $\angle AOB$ and $\overrightarrow{EB}$ is opposite to
$\overrightarrow{AC}$. Consequently, we do not get a physically
realizable value of $r_+$. The optimal $P_{POVM}$ strategy then  occurs at
$r_-=1$ and $r_+=0$ (i.e., $E$ coincides with $B$), one has
$P_{POVM}
=-\overrightarrow{OB} \cdot \overrightarrow{BD}= p_-
(1-|\alpha|^2)$, which is Eq. (\ref{case2}).

\section{discussion}


\begin{figure}
\includegraphics[width=8cm]{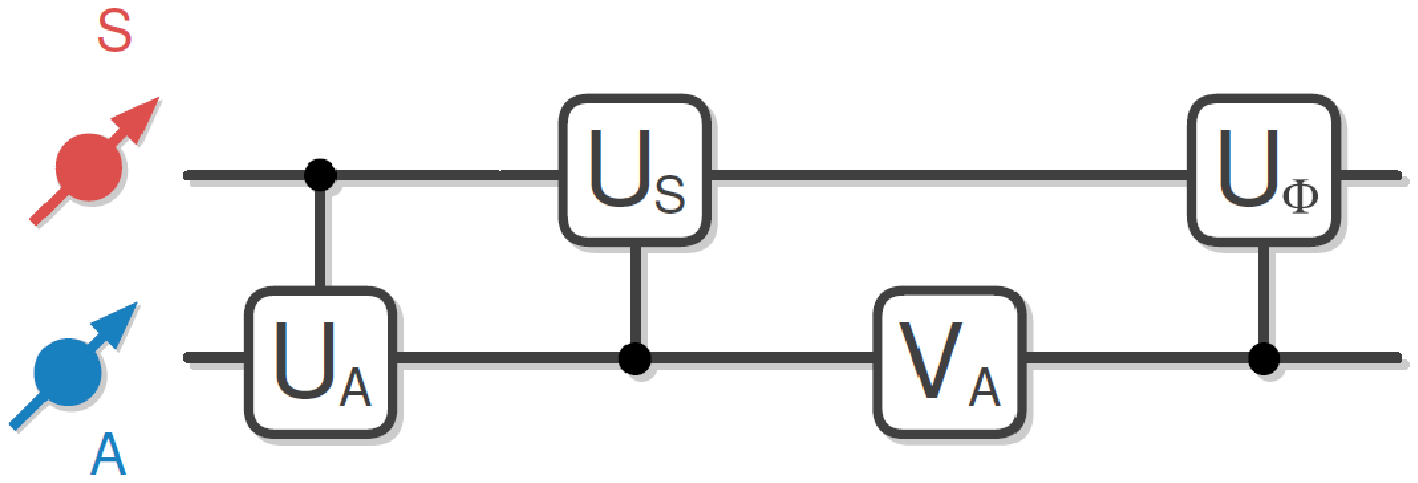} \\
 \caption{\textbf{Realization of the General Unitary Transformation.} For the initial states $|\psi_+\rangle=|0\rangle$, $|\psi_-\rangle=\sqrt{1-|\alpha|^2}|1\rangle + \alpha |0\rangle$ and $|k\rangle_a = |0\rangle_a$, the unitary transformation $\mathcal{U}$ in Eq. (\ref{Utrans}) can be realized in four steps:
(i) controlled-$U_A$ with the system $S$ being the control  qubit;
(ii) controlled-$U_S$ where the system $S$ is controlled by the ancilla $A$;
 (iii) local unitary $V_A$ on the auxiliary qubit;
 (iv) controlled-$U_{\Phi}$ with the same control qubit and target as the second step.
  The single qubit operations $U_A=|\phi_A\rangle_a \langle0|+|\bar{\phi}_A\rangle_a \langle1|$, $U_S=|\phi_S\rangle \langle0|+|\bar{\phi}_S\rangle \langle1|$, $V_A=|\phi_V\rangle_a \langle0|+|\bar{\phi}_V\rangle_a \langle1|$, and $U_{\Phi}=|\Phi\rangle \langle0|+|\bar{\Phi}\rangle \langle1|$, with $|\phi_A\rangle_a=(\sqrt{(1-|\alpha_+|^2)(1-|\alpha_-|^2)}|0\rangle_a+\sqrt{|\alpha_+|^2+|\alpha_-|^2-2|\alpha|^2}|1\rangle_a)/\sqrt{1-|\alpha|^2}$, $|\phi_S\rangle =(\sqrt{1-|\alpha_-|^2} \alpha_+^* |0\rangle+\sqrt{1-|\alpha_+|^2} \alpha_-^* |1\rangle )/\sqrt{|\alpha_+|^2+|\alpha_-|^2-2|\alpha|^2}$, and $|\phi_V\rangle_a=\sqrt{1-|\alpha_+|^2} |0\rangle_a + \alpha_+ |1\rangle_a$. Here, the states with a bar, $|\bar{\phi}_A\rangle_a$, $|\bar{\phi}_S\rangle $ , and $|\bar{\phi}_V\rangle_a$, denote $(i \sigma_{y,a} |\phi_A\rangle_a)^*$, $(-i \sigma_y |\phi_S\rangle)^*$, and $(i \sigma_{y,a}|\phi_V\rangle_a)^*$.
  }\label{fig4}
 \end{figure}

  \begin{figure}
\includegraphics[width=8cm]{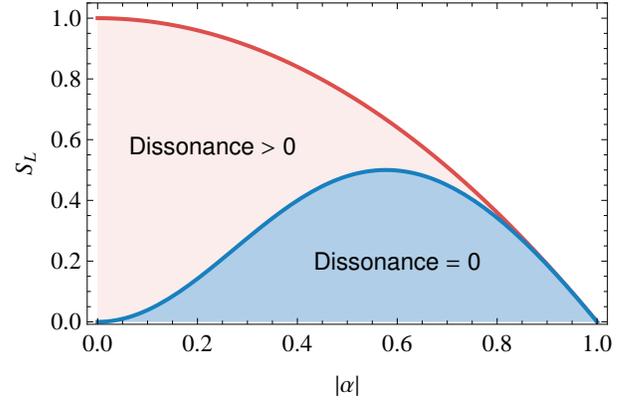} \\
 \caption{
\textbf{Necessary criterion for requiring dissonance in AOSD based
on linear entropy or purity.} The linear entropy reads
$\mathcal{S}_L(\rho) =2-2 \;\tr \rho^2 =4 p_+ p_- (1-|\alpha|^2) $,
and the purity $\mathcal{P}(\rho)=1-\mathcal{S}_L(\rho)$. For a
given amount of $|\alpha|$, when $\mathcal{S}_L(\rho)\leq
4(1-|\alpha|^2)|\alpha|^2 /(1+|\alpha|^2)^2$, the dissonance for the
AOSD is zero (see the region below the dashed line).
we note that $\mathcal{S}^{\rm max}_L(\rho)=1/2$ when $|\alpha|=1/\sqrt{3}$.
This means that if $\mathcal{S}_L(\rho)>1/2$, then the dissonance is
necessarily needed for the AOSD.} \label{fig5}
\end{figure}

In summary,  based on  a sufficiently general AOSD protocol, we found that the
entanglement between the principal qubit and the ancilla is
completely unnecessary.  Moreover,  this quantum entanglement
can be arbitrarily zero by adjusting the parameters in the joint unitary
 transformation without affecting the success probability. Theoretically, this fact
 clearly indicates that dissonance plays  a key role
 in assisted optimal state discrimination other than
 entanglement. Experimentally, the absence of entanglement can be more easily observed because there is no restriction on the {\it a priori} probabilities.
In Fig. \ref{fig4}, we present a realization of the unitary transformation $\mathcal{U}$ in Eq. (\ref{Utrans}) for the initial states $|\psi_+\rangle=|0\rangle$, $|\psi_-\rangle=\sqrt{1-|\alpha|^2}|1\rangle + \alpha |0\rangle$ and $|k\rangle_a = |0\rangle_a$ by using single-qubit gates and two-qubit controlled-unitary gates. These gates can be
demonstrated experimentally in many systems \cite{ChowSC,BrunnerQD} in recent years.
The success probability of state discrimination is determined by
 steps (i) to (iii), which transform the system-ancilla state into
 \begin{subequations}\label{Utrans1}
\begin{align}
|\psi_+\rangle|k\rangle_a\ \rightarrow \  \sqrt{1-|\alpha_+|^2}\;|0\rangle|0\rangle_a+\alpha_+|0\rangle|1\rangle_a, \\
|\psi_-\rangle|k\rangle_a\ \rightarrow \  \sqrt{1-|\alpha_-|^2}\;|1\rangle|0\rangle_a+\alpha_-|0\rangle|1\rangle_a.
\end{align}
\end{subequations}
It is not affected by the controlled-$U_{\Phi}$ in step (iv), which can adjust the correlations in state (\ref{rho}).




Let us also reiterate a necessary criterion for the requirement of dissonance in
AOSD based on linear entropy.  Under the general protocol,
Alice and Bob share the entangled state $|\zeta\rangle$, encoded in the
basis of the polarization of the qubit,
Bob can acquire knowledge of the linear entropy $\mathcal{S}_L(\rho)$ of Alice's qubit.
If   $\mathcal{S}_L(\rho)>1/2$, he can be sure that Alice needs
dissonance for her AOSD (see Fig. \ref{fig5}). Finally, we would
like to mention that local distinguishability of multipartite
orthogonal quantum states was studied in Ref. \cite{walg}  where again the
local discrimination of entangled states  does not require any
entanglement.

\vspace{3mm}

\indent{\bf Acknowledgements}
F.L.Z. is supported by NSF of China (Grant No. 11105097). J.L.C. is
supported by National Basic Research Program (973 Program) of China
under Grant No. 2012CB921900, NSF of China (Grant Nos. 10975075 and
11175089) and partly supported by National Research Foundation and
Ministry of Education of Singapore.

\vspace{3mm}

{\bf Author contributions} F.L.Z and J.L.C. initiated the idea.
F.L.Z. derived the formulas and prepared the figures. J.L.C., F.L.Z.
L.C.K. and V.V. wrote the main manuscript text. All authors
contributed to the derivation and the manuscript.

\vspace{3mm}

{\bf Additional information}

\textbf{Competing financial interests:} The authors declare no
competing financial interests.


\begin{thebibliography}{99}

\bibitem{Horodecki09} Horodecki, R., Horodecki, P., Horodecki, M. \& Horodecki, K. Quantum entanglement.  \emph{Rev. Mod. Phys.} \textbf{81}, 865-942 (2009).

\bibitem{Bell}
Bell, J. S. On the Einstein Podolsky Rosen paradox. \emph{Physics}
(Long Island City, N.Y.) \textbf{1}, 195-200 (1964).

\bibitem{PhysRevLett.88.017901}
Ollivier, H. \& Zurek, W. H. Quantum discord: a measure of the quantumness of correlations. \emph{Phys. Rev. Lett.} \textbf{88}, 017901
(2001).




\bibitem{henderson2001classical}
Henderson, L. \& Vedral V. Classical, quantum and total correlations. \emph{J. Phys. A} \textbf{34}, 6899-6905 (2001).



\bibitem{key}
Ekert, A. K. Quantum cryptography based on Bell's theorem. \emph{Phys. Rev. Lett.} \textbf{67}, 661-663 (1991).

\bibitem{lanyon2008experimental}
Lanyon, B. P., Barbieri, M., Almeida, M. P. \& White, A. G. Experimental quantum computing without entanglement. \emph{Phys. Rev. Lett.}
\textbf{101}, 200501 (2008).

\bibitem{datta2008quantum}
Datta, A., Shaji, A. \& Caves, C. M. Quantum discord and the power of one qubit. \emph{Phys. Rev. Lett.} \textbf{100},
050502 (2008).

\bibitem{cavalcanti2011operational}
Cavalcanti, D. et al. Operational interpretations of quantum discord. \emph{Phys. Rev. A} \textbf{83}, 032324 (2011).

\bibitem{madhok2011interpreting}
Madhok, V. \& Datta, A. Interpreting quantum discord through quantum state merging. \emph{Phys. Rev. A} \textbf{83}, 032323 (2011).

\bibitem{piani2011all}
Piani, M. et al. All nonclassical correlations can be activated into distillable entanglement. \emph{Phys. Rev. Lett.} \textbf{106}, 220403 (2011).

\bibitem{streltsov2011linking}
Streltsov, A., Kampermann, H. \& Bru{\ss}, D. Linking quantum discord to entanglement in a measurement. \emph{Phys. Rev. Lett.}
\textbf{106}, 160401 (2011).


\bibitem{modi2010unified}
Modi, K., Paterek, T., Son, W., Vedral, V. \& Williamson, M. Unified view of quantum and classical correlations. \emph{Phys.
Rev. Lett.} \textbf{104}, 080501 (2010).

\bibitem{roa2011dissonance}
Roa, L., Retamal, J. C. \& Alid-Vaccarezza, M. Dissonance is required for assisted optimal state discrimination. \emph{Phys. Rev. Lett.}
\textbf{107}, 080401 (2011).


\bibitem{john1996mathematical}
Neumann, J. V.  \emph{Mathematical Foundations of Quantum Mechanics},
Vol. \textbf{2} (Princeton University Press, 1996).


%


\bibitem{POVM2008}
Jafarizadeh, M. A., Rezaei, M., Karimi, N. \& Amiri, A. R. Optimal unambiguous discrimination of quantum states. \emph{Phys. Rev.
A} \textbf{77}, 042314 (2008).

%




\bibitem{Horo}
Horodecki, M., Horodecki, P.  \& Horodecki, R. General teleportation channel, singlet fraction, and quasidistillation. \emph{Phys. Rev. A}
\textbf{60}, 1888-1898 (1999).


\bibitem{Kim1}
Roa, L., Delgado, A. \& Fuentes-Guridi, I. Optimal conclusive
teleportation of quantum states. \emph{Phys. Rev. A} \textbf{68},
022310 (2003).

\bibitem{Kim2}
Kim, H., Cheong, Y. W. \& Lee, H. W. Generalized measurement and
conclusive teleportation with nonmaximal entanglement. \emph{Phys.
Rev. A} \textbf{70}, 012309 (2004).

\bibitem{Wootters98}
Wootters, W. K. Entanglement of formation of an arbitrary state of two qubits. \emph{Phys. Rev. Lett.} \textbf{80}, 2245-2248 (1998).

\bibitem{koashi2004monogamy}
Koashi, M. \& Winter, A. Monogamy of quantum entanglement and other correlations. \emph{Phys. Rev. A} \textbf{69}, 022309 (2004).


\bibitem{coff}
Coffman, V., Kundu, J. \& Wootters, W. K. Distributed entanglement. \emph{Phys. Rev. A} \textbf{61},
052306 (2000).

\bibitem{walg}
Walgate, J., Short, A. J., Hardy, L. \& Vedral, V. Local distinguishability of multipartite orthogonal quantum states. \emph{Phys. Rev. Lett.}
\textbf{85}, 4972-4975 (2000).

\bibitem{ChowSC}
Chow, J. M. et al. Universal quantum gate set approaching fault-tolerant thresholds with superconducting qubits. \emph{Phys. Rev. Lett.} \textbf{109}, 060501 (2012).

\bibitem{BrunnerQD}
Brunner, R. et al. Two-qubit gate of combined single-spin rotation and interdot spin exchange in a double quantum dot.  \emph{Phys. Rev. Lett.} \textbf{107}, 146801 (2011).

\end{thebibliography}
\end{document}